
\def\Z_2{{Z \!\!\! Z}_2}
\def\[{[\![}
\def\]{]\!]}

\baselineskip=18pt
\leftskip 36pt

\hfill INRNE-TH-94/4

\noindent
{\bf Algebraic structure of the Green's ansatz and its
$q-$deformed analogue}

\vskip 32pt
\noindent
T. D. Palev\footnote*{Permanent address: Institute for Nuclear Research
and Nuclear Energy, Boul. Tsarigradsko Chausse 72,
1784 Sofia, Bulgaria; E-mail
palev@bgearn.bitnet}

\noindent
Arnold Sommerfeld Institute for Mathematical Physics,
Technical University of Clausthal, 38678
Clausthal-Zellerfeld, Germany

\vskip 32pt
{\bf Abstract.} The algebraic structure of the Green's
ansatz is analyzed in such a way that its generalization to
the case of $q-$deformed para-Bose and para-Fermi operators
is becoming evident. To this end the underlying Lie
(super)algebraic properties of the parastatistics are
essentially used.

\vskip 48pt
Already in his first paper on parastatistics [1] Green
developed a technique, the Green's ansatz technique [2],
appropriate for constructing new (reducible)
representations for any set of para-Bose (pB) or para-Fermi
(pF) operators. Although quite clear as a mathematical
device, the inner structure of the Green's ansatz remains
somehow not completely understood up to now. Just to give
an example, consider $m$ pairs of para-Bose creation and
annihilation operators $b_r^\pm (p), \; r=1,\ldots ,m$ of
order $p$.  Then each such operator is represented by
a sum

$$b_r^\pm(p) =\sum_{k=1}^p b_r^{\pm k}, \eqno(1) $$

\noindent
where for any value of $k$ the $b_r^{\pm k}$ operators
obey the Bose commutation relations (here and throughout
$[x,y]=xy-yx,\;\{x,y\}=xy+yx$)

$$[b_r^{-k},b_s^k]=\delta _{rs},\quad
[b_r^{-k},b_s^{-k}]=[b_r^{k},b_s^k]=0,\quad
\forall \; r,s,\eqno(2)$$

\noindent
and for $i\neq j$ all operators anticommute,

$$\{b_r^{\xi i},b_s^{\eta j}\}=0,\quad \forall  \; \xi , \eta =\pm ,
\quad i \neq j,\quad r,s. \eqno (3)  $$

\noindent
One natural question that arises in relation to the above
construction is why the Bose operators partially commute
and partially anticommute. Is there any deeper reason
behind?  The purpose of the present note is to answer
questions like that and in fact to show that the Green's
ansatz construction (1)-(3) is a very natural one from a
Lie superalgebraic point of view.

Much of the motivation for the present work stems from the
recent interest in  deformed para-Bose  and para-Fermi
operators from various points of view: deformed
paraoscillators [3-10] and, more generally, deformed
oscillators (see [11, 12] also for collection of references
in this respect), supersingleton Fock representations of
$U_q[osp(1/4)]$ and its singleton structure [13],
integrable systems [14-18] and $q-$parasuperalgebras [19].

In the applications we mentioned above one of the questions
is how to construct representations of the deformed
paraoperators.  In the nondeformed case the Green's ansatz
gives in principle an answer to this question.  Therefore it is
natural to try to extent the same technique to the deformed case.
In the present paper we will analyze the algebraic structure of
the Green's ansatz in such a way that its generalization to the
quantum case will become evident.  To this end we essentially use
the circumstance (see also Corollary 1) that any $n$ pairs
$F_1^\pm,\ldots , F_n^\pm$ of pF operators generate the simple
Lie algebra $so(2n+1)$ [20, 21], whereas $m$ pairs of pB
operators $B_1^\pm,\ldots , B_m^\pm$ generate a Lie superalgebra
[22], which is isomorphic  to the basic Lie superalgebra
$osp(1/2m)$  [23], denoted also as $B(0/m)$ [24].

In order to be slightly more general and to treat the pB
and the pF operators simultaneously, denote by $G(n/m)$ a
$2(m+n)$-dimensional $\Z_2$-graded linear space ($\Z_2
\equiv (0,1)$) with a basis as follows:
$$\eqalignno{
& {\rm even \; basis \; vectors } \quad C_j^\pm(0)\equiv F_j^\pm
,\;j=1,\ldots,n, & (4)\cr
& {\rm odd \; basis \; vectors } \quad C_i^\pm(1)\equiv B_i^\pm
,\;i=1,\ldots ,m. & (5) \cr
}$$

\noindent
Let $U(n/m)$ be the free associative unital (= with
unity) superalgebra  with generators (4) and (5), grading
induced from the grading of the generators and relations

$$\[\[C_i^\xi (\alpha),C_j^\eta (\beta) \],C_k^\varepsilon (\gamma) \]=
2\varepsilon^{\gamma} \delta_{\beta \gamma}\delta_{jk}
\delta_{\varepsilon,-\eta}C_i^\xi (\alpha)-
2\varepsilon^\gamma (-1)^{\beta \gamma}\delta_{\alpha \gamma}
\delta_{ik}\delta_{\varepsilon,-\xi}
C_j^\eta (\beta ),  \eqno(6)  $$

\noindent
where
$\quad \xi ,\eta ,\varepsilon =\pm ,\quad\alpha ,\beta
,\gamma\in \Z_2$
and $i,j,k $ take all possible values according to (4) and
(5). In (6) and throughout $\[\;,\;\]$ is a
supercommutator, defined on any two homogeneous elements
$a,b$ from $U(n/m)$ as

$$\[a,b\]=ab-(-1)^{deg(a)deg(b)}ba. \eqno(7)  $$

In the case $\alpha =\beta =\gamma =0$ (6) reduces to

$$[[F_i^\xi ,F_j^\eta ],F_k^\varepsilon ]=
2\delta_{jk}
\delta_{\varepsilon,-\eta}F_i^\xi -
2\delta_{ik}\delta_{\varepsilon,-\xi}
F_j^\eta ,  \eqno(8)  $$

\noindent
whereas for $\alpha =\beta =\gamma =1$ it gives

$$[\{B_i^\xi ,B_j^\eta  \},B_k^\varepsilon ]=
2\varepsilon \delta_{jk}
\delta_{\varepsilon,-\eta}B_i^\xi +
2\varepsilon
\delta_{ik}\delta_{\varepsilon,-\xi}
B_j^\eta ,  \eqno(9)  $$

Equations (8) and (9) are the defining relations for the
para-Fermi and para-Bose operators, respectively [1].

The relations (6) define a structure of a Lie-super triple
system [25] on $G(n/m)\subset U(n/m)$ with a triple product
$G(n/m)\otimes G(n/m)\otimes G(n/m) \rightarrow G(n/m)$
defined as

$$\[\[x,y\],z\]=2<y \vert z>x-2(-1)^{deg(x)deg(y)}<x \vert
z>y \in G(n/m),\; \forall x,y,z \in G(n/m), \eqno(10) $$

\noindent
where the bilinear form  $<x \vert y>$ is defined in
agreement with (6) to be [25]

$$<C_i^\xi(\alpha) \vert C_j^\eta(\beta) >=\eta^\alpha
\delta_{ij}\delta_{\alpha \beta} \delta_{\xi,-\eta},
\quad \xi, \eta =\pm, \quad \alpha , \beta
\in \Z_2 . \eqno(11)  $$

Consider $U(n/m)$ as a Lie superalgebra (LS) with a
supercommutator (7). Then it is straightforward to check
that ($lin.env.$ = linear envelope)

$$B(n/m)=lin.env.\{\[C_i^\xi (\alpha),C_j^\eta (\beta)\],\;
C_k^\varepsilon (\gamma)|\forall i,j,k,\;
\xi ,\eta ,\varepsilon =\pm ,\; \alpha ,\beta
,\gamma\in \Z_2 \} \eqno(12)  $$

\noindent
is a subalgebra of the LS $U(n/m)$.
\smallskip
{\it Proposition 1} [26]. The LS $B(n/m)$ is isomorphic to
the orthosymplectic LS $osp(2n+1/2m)$. The associative
superalgebra $U(n/m)$ is its universal enveloping algebra
$U[osp(2n+1/2m)]$.

\smallskip
The first part of the proposition was proved in [26]. The
second part follows from the following two
observations:

(i) The supercommutation relations between all generators
of $osp(2n+1/2m)$ (which constitute a basis in the
underlying linear space) follow from the relations (6)
between only the Lie-super triple generators (4) and (5).

(ii) The universal enveloping algebra of a given LS is the
free associative unital algebra of its generators and the
supercommutation relations they satisfy.

Observe that everywhere in the above considerations the
para-Fermi operators appear as even (i.e.  bosonic)
variables, whereas the parabosons are odd (i.e.  fermionic)
operators. Moreover the parabosons do not commute with the
parafermions. To the same conclusion arrived recently also
Okubo [25] and Macfarlane [15].

As an immediate consequence of the above proposition we have

\smallskip
{\it Corollary 1}.

\noindent
(a) [20, 21] The free associative unital algebra of the pF
operators (4) is isomorphic to the universal enveloping
algebra $U[so(2n+1)]$ of the orthogonal Lie algebra
$so(2n+1)$;

$$so(2n+1)=lin.env.\{[F_i^\xi ,F_j^\eta ],F_k^\varepsilon
|i,j,k=1,\ldots,n;\;\xi ,\eta ,\varepsilon =\pm \}
\subset U[so(2n+1)];\eqno(13)  $$

\noindent
(b) [23] The free associative unital algebra of the pB
operators (5) is isomorphic to the universal enveloping
algebra $U[osp(1/2m]$ of the orthosymplectic Lie
superalgebra $osp(1/2m)$;

$$osp(1/2m)=lin.env.\{ \{B_i^\xi ,B_j^\eta \},B_k^\varepsilon
|i,j,k=1,\ldots,m;\;\xi ,\eta ,\varepsilon =\pm \}
\subset U[osp(1/2m)].\eqno(14)  $$

\smallskip
{\it Corollary 2.} The representation theory of the
Lie-super triple system $G(n/m)$ with generators (4), (5)
and relations (6) is completely equivalent to the
representation theory of the orthosymplectic LS
$osp(2n+1/2m)$. In particular, the problem to construct the
representations of $n$ pairs of pF operators (4) is
equivalent to the problem to construct the representations
of the Lie algebra $so(2n+1)$; similarly, the
representation theory of $m$ pairs of pB operators is the
same as the representation theory of the LS $osp(1/2m)$.

\smallskip
The finite-dimensional representations of $so(2n+1)$ are
known, they have been explicitly constructed [27]. All
representations of the pF operators corresponding to a
fixed order of the parastatistics are among the
finite-dimensional representations. The pF operators have
however several other representations [28], including
representations with degenerate vacua. In a more practical
aspect the results of [27] are unfortunately not so useful
for the para-Fermi statistics. The point is that the
transformation relations of the Gel'fand-Zetlin basis in [27] are
given for a set of $2n$ operators (generating the rest of all
$2n^2+n$ generators), which are different from the pF operators
and also different from the $3n$ Chevalley generators. The
relations between the pF operators and the operators used in [27]
are not linear.

The above corollaries are not of big practical use also
for the representations of the para-Bose operators or, more
generally, of the Lie-super triple system $G(n/m)$.  The
finite-dimensional representations of $osp(2n+1/2m)$ have
been classified only so far [29].  Explicit expressions for
the matrix elements are available only for low rank algebras
(see [30] and the references therein). Moreover the
interesting representations of the pB operators are
infinite-dimensional. The Lie-super triple system $G(n/m)$
and hence $osp(2n+1/2m)$ have however one simple but
important representation, the Fock representation, which is
of particular interest for our considerations.

\smallskip
{\it Proposition 2} [26]. Denote by $W(n/m)$ the
antisymmetric Clifford-Weyl superalgebra, namely the associative
algebra generated by $n$ pairs of Fermi creation and
annihilation operators (CAOs) $f_i^\pm \equiv c_i^\pm(0)$,
$i,j=1,\ldots ,n$,

$$\{f_i^\xi,f_j^\eta\}\equiv \{c_i^\xi(0),c_j^\eta(0) \}=
\delta_{ij}\delta_{\xi,-\eta},\quad
\xi ,\eta  =\pm \eqno(15)  $$

\noindent
and $m$ pairs of Bose CAOs $b_j^\pm \equiv c_j^\pm(1)$,
$i,j=1,\ldots ,m$,

$$[b_i^\xi,b_j^\eta]\equiv [c_i^\xi(1),c_j^\eta(1)]=
\eta \delta_{ij}\delta_{\xi,-\eta}, \quad
\xi ,\eta  =\pm \eqno(16)   $$

\noindent
under the condition that the Bose operators anticommute
with the Fermi operators,

$$\{f_i^\xi,b_j^\eta\}\equiv \{c_i^\xi(0),c_j^\eta(1) \}=0,
\quad \xi ,\eta=\pm\quad, \quad i=1,\ldots,n,
\quad\;j=1,\ldots,m. \eqno(17)  $$

\noindent
$W(n/m)$ is  an associative superalgebra with grading
induced from the requirement that the Fermi operators are
even elements and the Bose operators are odd. Consider
$W(n/m)$ as an algebra of (linear) operators in the
corresponding Fock space $H\equiv H(n/m)$, $W(n/m)\subset End(H)$.
Then the map

$$\pi :osp(2n+1/2m)\rightarrow W(n/m)\quad {\rm defined \; as}
\quad  \pi(C_i^\xi(\alpha))=c_i^\xi(\alpha),
\quad \forall \; \xi=\pm \; {\rm and}\; i \eqno(18) $$

\noindent
is a representation, a Fock representation, of
$osp(2n+1/2m)$ or, which is the same, a representation of
the Lie-super triple system $G(n/m)$.

In order to prove the proposition one has simply to check
that the relations (6) remain valid after the replacement
$C_i^\xi(\alpha) \;\rightarrow \;  c_i^\xi(\alpha)$.

In case of $so(2n+1)$ or equivalently of $n$ pairs of
pF operators (resp. of $osp(1/2m)$ or equivalently of
$m$ pairs of pB operators) the proposition 2 reduces to the
usual representation of the pF operators with Fermi operators
(resp. of the pB operators with Bose operators).

The relevant for us conclusion is that the operators (4)
and (5) generate an associative  superalgebra, namely
$U[osp(2n+1/2m)]$ (Proposition 1) and that we know at least
one representation of $U[osp(2n+1/2m)]$, namely its Fock
representation (Proposition 2).

Set for simplicity $L=osp(2n+1/2m)$, $U=U[osp(2n+1/2m)]$
and let $L^{\otimes p}$ and
$U^{\otimes p}$ be their $p^{th}$ tensorial powers.
Introduce the following notation ($e$ is the unity of $U$):

$$L^k=\{e_1\otimes \ldots e_{k-1}\otimes a \otimes
e_{k+1}\otimes \ldots \otimes e_p| a\in L,\;e_i=e \; \forall
\; i\neq k \}.\eqno(19) $$

$$U^k=\{e_1\otimes \ldots e_{k-1}\otimes u \otimes
e_{k+1}\otimes \ldots \otimes e_p|u\in U,\;  e_i=e \; \forall
\; i\neq k \}.\eqno(20) $$

\noindent
Then the map $\tau^k: L\rightarrow L^k\subset U^k $

$$\tau^k(a)= e_1\otimes \ldots e_{k-1}\otimes a \otimes
e_{k+1}\otimes \ldots \otimes e_p, \quad
a\in L,\; e_i=e \; \forall
\; i\neq k. \eqno(21)  $$

\noindent
is a Lie superalgebra morphism of $L$ onto $L^k$; the
same map (21) considered for all $a\in U$ is an associative algebra
morphism of $U$ onto $U^k$.

The set of the elements (21) generate $U^k$ and, since

$$U^{\otimes p}=U^1 U^2\ldots U^p, \eqno(22)$$

\noindent
the elements (21) considered for all $k=1,\ldots,p$ generate
$U^{\otimes p}$.

The sum

$$\Delta^{(p)}=\tau^1+\tau^2+\ldots+\tau^p\; : \;
L \longrightarrow  L^{\otimes p} \eqno(23)$$

\noindent
is a Lie superalgebra morphism, the "diagonal" LS morphism,
of $L$ into $L^{\otimes p}$, which is extended to a
morphism of the associative algebra $U$ into the
associative algebra $U^{\otimes p}$ in a natural way:

$$\Delta^{(p)}(a_1a_2\ldots a_m)=
\Delta^{(p)}(a_1)\Delta^{(p)}(a_2)\ldots\Delta^{(p)}(a_m)
\quad \forall \; a_1,a_2,\ldots, a_m \in L.\eqno(24) $$

Let $\pi^1,\pi^2,\ldots,\pi^p$ be (not necessarily different)
representations of $L=osp(2n+1/2m)$ (and hence of
$U=U[osp(2n+1/2m)]$) in the $\Z_2$-graded linear
spaces $H^1,H^2,\ldots,H^p $, respectively, i.e., the
operators $\pi^k[C_r^\pm(\alpha)] \in
End(H^k)$, $ k=1,\ldots,p$, satisfy the Lie-super triple
relations (6) and

$$deg\{\pi^k[C_r^\pm(\alpha)] \}=\alpha .\eqno(25) $$

\noindent
Then

$$ \pi^1\otimes\pi^2\otimes\ldots\otimes\pi^p:
U^{\otimes p} \longrightarrow
End(H^1\otimes H^2\otimes\ldots\otimes H^p) \eqno(26)$$

\noindent
gives a representation of both the LS $L^{\otimes p}$ and of the
associative algebra $U^{\otimes p}$.

\noindent
The composition maps ($ k=1,\ldots,p$)

$$(\pi^1\otimes\pi^2\otimes\ldots\otimes\pi^p)\circ\tau^k:
U[osp(2n+1/2m)] \longrightarrow End(H^1\otimes
H^2\otimes\ldots\otimes H^p), \eqno(27)$$
$$(\pi^1\otimes\pi^2\otimes\ldots\otimes\pi^p)\circ\Delta^{(p)}:
U[osp(2n+1/2m] \longrightarrow End(H^1\otimes
H^2\otimes\ldots\otimes H^p)
\eqno(28)$$

\noindent
give representations of both the LS $osp(2n+1/2m)$ and the
associative algebra $U[osp(2n+1/2m)]$. Therefore the operators

$$
\eqalignno{
& {\hat c}_r^{\pm k}(\alpha)=
  [(\pi^1\otimes\pi^2\otimes\ldots\otimes\pi^p)
  \circ\tau^k]C_r^\pm (\alpha)
  =id^1\otimes\ldots\otimes id^{k-1}\otimes\pi^k[C_r^\pm (\alpha)]
  \otimes id^{k+1}\otimes\ldots\otimes id^p \cr
& \in End(H^1\otimes H^2\otimes\ldots\otimes H^p),\quad
  k=1,\ldots,p,\quad \alpha =\pm,  & (29) \cr
}$$
and
$${\hat c}_r^\pm (p,\alpha)=
  [(\pi^1\otimes\pi^2\otimes\ldots\otimes\pi^p)
  \circ\Delta^{(p)} ]C_r^\pm (\alpha)=
  \sum_{k=1}^p {\hat c}_r^{\pm k}(\alpha) \eqno(30)   $$

\noindent
satisfy the Lie-super triple relations (6). From the very
definition of a tensor product of associative algebras [31]
we obtain (for all $r,s$ according to (4) and (5)
and $\xi,\;\eta=\pm$):

$$\[{\hat c}_r^{\xi i}(\alpha),{\hat c}_s^{\eta j}(\beta)\]
\equiv {\hat c}_r^{\xi i}(\alpha){\hat c}_s^{\eta j}(\beta)
-(-1)^{\alpha\beta}
{\hat c}_s^{\eta j}(\beta){\hat c}_r^{\xi i}(\alpha)=0,
\quad i\neq j=1,\ldots,p.\eqno(31)$$

In particular for $\alpha=1$

$${\hat b}_r^{\pm k}\equiv {\hat c}_r^{\pm k}(1)
  =id^1\otimes\ldots\otimes id^{k-1}\otimes\pi^k(B_r^\pm )
  \otimes id^{k+1}\otimes\ldots\otimes id^p, \eqno(32)$$

$${\hat b}_r^\pm(p) \equiv {\hat c}_r^\pm (p,1)
  = \sum_{k=1}^p {\hat b}_r^{\pm k},\quad
  r=1,\ldots,m,\eqno(33) $$

\noindent
and

$$\{{\hat b}_r^{\xi i},{\hat b}_s^{\eta j} \}=0
\quad i\neq j=1,\ldots,p,\quad r,s=1,\ldots,m,
\quad \xi,\eta=\pm,\eqno(34)$$

\noindent
whereas  the operators ${\hat b}_r^{\pm k}$ with the same
upper case index $k$ satisfy the pB relations (9) and may
be also other, particular for the representation $\pi^k$,
relations.

Similarly for $\alpha =0$

$${\hat f}_r^{\pm k}\equiv {\hat c}_r^{\pm k}(0)
  =id^1\otimes\ldots\otimes id^{k-1}\otimes\pi^k(F_r^\pm )
  \otimes id^{k+1}\otimes\ldots\otimes id^p,  \eqno(35)$$

$${\hat f}_r^\pm(p) \equiv {\hat c}_r^\pm (p,0)
  = \sum_{k=1}^p {\hat f}_r^{\pm k},\quad
  r=1,\ldots,n,\eqno(36) $$

\noindent
and

$$[{\hat f}_r^{\xi i},{\hat f}_s^{\eta j}]=0
\quad i\neq j=1,\ldots,p,\quad r,s=1,\ldots,n,
\quad \xi,\eta=\pm. \eqno(37)$$

Consider now  the important  case when all representations
$\pi^1,\pi^2,\ldots,\pi^p$ are the same and coincide with
the Fock representation, namely $\pi$ is a morphism of
$U[osp(2n+1/2m)]$ onto the Clifford-Weyl algebra $W(n/m)$
defined in (18),

$$\pi^1=\pi^2=\ldots=\pi^p=\pi.\eqno(38)$$

\noindent
In order to distinguish this particular case we do not
write any more hats over the operators. Then from (30)  we
obtain

$$c_r^\pm (p,\alpha)=
  [\pi^{\otimes p} \circ\Delta^{(p)}]C_r^\pm (\alpha)=
  \sum_{k=1}^p c_r^{\pm k}(\alpha)\in
  End(H^{\otimes p}), \eqno(39)   $$

\noindent
where according to (29) and (18)

$$c_r^{\pm k}(\alpha)
  =id^1\otimes\ldots\otimes id^{k-1}\otimes c_r^\pm (\alpha)
  \otimes id^{k+1}\otimes\ldots\otimes id^p ,\quad
  k=1,\ldots,p,\quad \alpha =\pm, \eqno(40)$$

\noindent
and the operators   $ c_r^{\pm k}(\alpha)$ satisfy
according to (15) - (17) and (31) the relations (see (7)):

$$\[c_r^{\xi i}(\alpha),c_s^{\eta j}(\beta) \]=\eta^\alpha
\delta_{rs}\delta_{ij}\delta_{\alpha \beta} \delta_{\xi,-\eta},
\quad \xi, \eta =\pm, \quad \alpha , \beta
\in \Z_2 . \eqno(41) $$

Setting $b_r^\pm(p)=c_r^\pm(p,1)$ we obtain from (39) and (41)
the Green's ansatz for the para-Bose operators of order $p$,

$$b_r^\pm (p)=
  [\pi^{\otimes p} \circ\Delta^{(p)}]B_r^\pm =
  \sum_{k=1}^p b_r^{\pm k}, \eqno(42) $$

 \noindent
 where

$$b_r^{\pm k}
  =id^1\otimes\ldots\otimes id^{k-1}\otimes b_r^\pm
  \otimes id^{k+1}\otimes\ldots\otimes id^p ,\quad
  k=1,\ldots,p,\quad r=1,\ldots,m. \eqno(43)$$

\noindent
As it follows from (40) (or immediately from (43), taking
into account that $b_r^\pm$ are odd operators) the Bose
operators $b_r^{\pm k} $ partially commute and partially
anticommute. More precisely,

$$[b_r^{-k},b_s^k]=\delta _{rs},\quad
[b_r^{-k},b_s^{-k}]=[b_r^{k},b_s^k]=0,\quad \forall \; k,r,s
\eqno(44)$$

\noindent
and for $i\neq j$ all operators anticommute,

$$\{b_r^{\xi i},b_s^{\eta j}\}=0, \quad \xi , \eta =\pm ,
\quad i \neq j. \eqno (45)  $$

Similarly setting $f_r^\pm(p)=c_r^\pm(p,0)$ we obtain from
(39) and (41) the Green's ansatz for the para-Fermi
operators of order $p$,

$$f_r^\pm (p)=
  [\pi^{\otimes p} \circ\Delta^{(p)}]F_r^\pm =
  \sum_{k=1}^p f_r^{\pm k}, \eqno(46) $$

 \noindent
 where

$$f_r^{\pm k}
  =id^1\otimes\ldots\otimes id^{k-1}\otimes f_r^\pm
  \otimes id^{k+1}\otimes\ldots\otimes id^p ,\quad
  k=1,\ldots,p,\quad r=1,\ldots,n, \eqno(47)$$

\noindent
Setting in (40) $\alpha=0$ (or directly from (47), taking into
account that $f_r^\pm$ are even operators) one obtains:

$$\{f_r^{-k},f_s^k\}=\delta _{rs},\quad
\{f_r^{-k},f_s^{-k}\}=\{f_r^{k},f_s^k\}=0,\quad \forall \;k,r,s,
\eqno(48)$$

\noindent
and for $i\neq j$ all operators commute,

$$[f_r^{\xi i},f_s^{\eta j}]=0, \quad \xi , \eta =\pm ,
\quad i \neq j. \eqno (49)  $$

{}From the above considerations it is clear that the Green's
ansatz representation (46) of the pF operators of order $p$
is simply given as a representation of the pF operators
$f_r^\pm(p)$, considered as generators of the universal
enveloping algebra $U[so(2n+1)]$), in the tensor product of
$p$ copies of (irreducible, finite-dimensional) Fock
representations of the Lie algebra $so(2n+1)$.

Similarly, the Green's ansatz representation (42) of the pB
operators of order $p$ gives a representation of the pB
operators  in the tensor product of $p$ copies of
(irreducible, infinite-dimensional) Fock representations of
the Lie superalgebra $osp(1/2m)$.

The eqs.(39), (41) generalize the concept of a Green's ansatz to
the case of Lie-super triple operators (4) and (5), which are
free generators with relations (6) of the universal enveloping
algebra $U[osp(2n+1/2m)]$.  The representation of the generators
$C_r^\pm(\alpha)$ (= the representation of $osp(2n+1/2m)$) is
realized in the tensor product space $H^{\otimes p}$ of $p$
copies of Fock representations (18) of $osp(2n+1/2m)$. Therefore
the Green's ansatz gives highly reducible representation of the
Lie-super triple operators. In particular this is the case if
only pB or pF operators are present.  If $|0>\in H$ is the
highest weight vector in $H$, then the irreducible subspace,
containing $|0>^{\otimes p}\in H^{\otimes p}$ carries a
representation, corresponding to an order of statistics $p$. The
other irreducible components of $H^{\otimes p}$ contain also
vacuum like states and among them are the highest weight vectors.
The corresponding representations however do not correspond
anymore to those with a fixed order of the statistics, namely to
representations with a unique vacuum states (see for example
[28]).  The problem to decompose $H^{\otimes p}$ into a direct
sum of irreducible subspaces with respect to the paraoperators (=
with respect to $osp(2n+1/2m)$) or even the simpler problem - to
extract the irreducible submodule, carrying only the
representation with an order of statistics $p$ - has not been
solved so far. The problem was not solved also for the case of
only pF operators ($m=0$) or only pB operators ($n=0$).

Passing to a short discussion of a possible generalization of the
Green's ansatz (39) to the case of deformed operators, we first
observe that in all cases the Green's ansatz is obtained (see
(39), (42) and (46)) as a two step procedure, namely as a
composition of two (associative algebra)  morphisms:

$$\eqalignno{
& \Delta^{(p)} :U[osp(2n+1/2m)] \longrightarrow
U[osp(2n+1/2m)]^{\otimes p}&\cr
& \pi^{\otimes p}: U[osp(2n+1/2m)]^{\otimes p}
\longrightarrow  End(H^{\otimes p}).  &(50)\cr
}$$

\noindent
In the following we  consider only such (one parameter)
deformations of the Lie-super triple generators (4) and (5),
which generate a Hopf deformation $U_q[osp(2n+1/2m)]$ of
$U[osp(2n+1/2m)]$.  By a Hopf deformation we mean a deformation
of $U[osp(2n+1/2m)]$, which preserves its Hopf algebra structure
(as defined, for instance, in [32]).

Denote by

$$\eqalignno{
& C_j^\pm(0)_q\equiv F_{jq}^\pm,\;j=1,\ldots,n, & (51)\cr
& C_i^\pm(1)_q\equiv B_{iq}^\pm,\;i=1,\ldots ,m. & (52) \cr
}$$

\noindent
a set of $2(n+m)$ deformed Lie-super triple generators, which at
$q \rightarrow 1$ reduce to (4) and (5). Suppose that they
satisfy one or more defining relations

$$\phi_q(C_1^\pm(0)_q,\ldots,C_n^\pm(0)_q,
   C_1^\pm(1)_q,\ldots,C_m^\pm(1)=0, \eqno(53)  $$

\noindent
which in the case $q \rightarrow 1$ reduce to the Lie-super
triple relations (6). According to what we have said above,
we require that the free unital associative algebra of the
generators (51) and (52) with relations (53) is a Hopf
deformation $U_q[osp(2n+1/2m)]$ of the universal enveloping
algebra  $U[osp(2n+1/2m)]$ of $osp(2n+1/2m)$. Such
operators do exist.  The operators generating $U_q[osp(3/2)]$
(the case $n=m=1$) were constructed in [33]. The Hopf deformation
of one pair of pB operators was carried out in [5], of two pairs
- in [34], of any number of pB operators - in [8-10].  The
deformation of any number of pF operators was obtained in [35].

The deformed version of the Green's ansatz of order $p$ we
are going to present is be based on the relation (see (39))

$$c_r^\pm (p,\alpha)_q=
  [\pi^{\otimes p} \circ\Delta^{(p)}]C_r^\pm (\alpha)_q.
  \eqno(54)   $$

In order to define a deformed analogue of the operator
$\Delta^{(p)}$ we use the circumstance that the superalgebra
$U=U[osp(2n+1/2m]$ is a Hopf superalgebra with a
comultiplication  $\Delta$, defined as

$$\Delta(a)=a\otimes e+e\otimes a,\quad \forall a\in L,
\quad \Delta(e)=e\otimes e. \eqno(55)$$

\noindent
{}From (21) and (23) we deduce that

$$\Delta^{(2)}=\Delta,\quad \Delta^{(3)}=
(id\otimes\Delta)\circ\Delta^{(2)},\quad
\Delta^{(k)}=[(id^{\otimes (k-2)})\otimes \Delta]\circ
\Delta^{(k-1)}.\eqno(56) $$

\noindent
The important point is that the operators $\Delta^{(k)}$
preserve the property to be morphisms also after the
quantization (=Hopf deformation) of $U(L)$, i.e., the map

$$\Delta^{(p)}:U_q[osp(2n+1/2m)] \longrightarrow
U_q[osp(2n+1/2m)]^{\otimes p}
\eqno(57)$$

\noindent
is an associative algebra morphism. Certainly in the
deformed case the eq. (55) has to be replaced with the
corresponding expression for the comultiplication on
$U_q[osp(2n+1/2m)]$.

In order to determine the deformed analogue of the operator
$\pi^{\otimes p}$ we first observe that if
$\pi :U_q[osp(2n+1/2m)]\rightarrow End(H) $ is a
representation of $U_q[osp(2n+1/2m)]$ in the linear space
$H$, then

$$\pi^{\otimes p}:U_q[osp(2n+1/2m)]^{\otimes p}
\longrightarrow End(H^{\otimes p}) \eqno(58) $$

\noindent
is a representation of $U_q[osp(2n+1/2m)]^{\otimes p}$.
In the nondeformed case $\pi$ is a morphism of
$U[osp(2n+1/2m)$ onto the Clifford-Weyl algebra $W(n/m)$
(see (18)). Therefore it is natural to assume that in the
deformed case $\pi$ is a morphism of $U_q[osp(2n+1/2m)]$
onto a deformed  algebra $W_q(n/m)$. The deformed
Clifford-Weyl algebra was defined in [33]. In case $n=0$
$W_q(0/m)$ is the associative superalgebra, generated by
$m$ triples $b_r^\pm,\; k_r=q^{N_r},\;r=1,\ldots,m$ of
commuting deformed Bose operators as defined in [36-38].
The morphism $\pi$ of $U_q[osp(1/2m)]$ onto $W_q(0/m)$,
namely the operators $\pi(B_{iq}^\pm)\in W_q(0/m)$
were constructed in [9]. In case $m=0$ $W_q(n/0)$ is the
associative algebra of $3n$ triples deformed Fermi operators
[39].  The morphism $\pi$ of $U_q[so(2n+1)]$ onto $W_q(n/0)$ is
given in [35]. Thus, we are ready to state the following result.

{\it Proposition 3.} If $\pi$ is the Fock representation of
$U_q[osp(1/2m)$ [9] (resp. of $U_q[so(2n+1)]$ [39] and
$\Delta^{(p)}$ is the operator (57) then eq. (54) defines the
deformed analogue of the Green ansatz of order $p$ for $m$ pairs
of deformed para-Bose operators (resp. for $n$ pairs of deformed
para-Fermi operators).

The proof is evident since the composition of the morphisms
$\pi^{\otimes p}$ and $\Delta^{(p)}$ is also a morphism. Hence
the operators $c_r^\pm (p,\alpha)_q$, defined in (54), satisfy
the same relations as $C_r^\pm (\alpha)_q$; in the limit $q
\rightarrow 1$ they reduce to the corresponding nondeformed
para-Bose or para-Fermi operators of order $p$.

The proposition 3 could be extended also to deformed Lie-super
triple systems. So far however such a deformation was carried out
only for the case $n=m=1$ [33].

As an example we write down the Green's ansatz of order $p=2$
related to $U_q[osp(1/4)$, namely the ansatz corresponding to two
pairs of deformed para-Bose operators $b_1^\pm(2)_q$ and
$b_2^\pm(2)_q$ [34]:

$$\eqalignno{
& b_1^+(2)_q=b_1^+ \otimes q^{N_1-2N_2-1/2} + q^{-N_1-2N_2-3/2}
\otimes b_1^+ +(q^{1/2}-q^{-7/2})b_2^+q^{-N_1-N_2} \otimes
b_1^+b_2^-q^{-N_2}, & (59a)\cr
& b_1^-(2)_q=b_1^-  \otimes q^{N_1+2N_2+3/2} +
q^{-N_1+2N_2+1/2} \otimes b_1^- +(q^{-1/2}-q^{7/2})b_1^-b_2^+q^{N_2}
\otimes b_2^-q^{N_1+N_2}, & (59b)\cr
& b_2^\xi(2)_q=b_2^\xi \otimes q^{N_2+1/2}+q^{-N_2-1/2}
\otimes b_2^\xi, \hskip 6pt \xi=\pm. & (60) \cr
}$$

This example indicates that the structure of the deformed Green's
ansatz is more involved.  There is in particular a big asymmetry
between the first pair of operators $b_1^\pm(2)_q$ and the second
pair $b_2^\pm(2)_q$. As a result the relatively simple problem to
decompose the tensor product of two Fock representation into
a direct sum of irreducible representations of $osp(1/4)$ [40]
becomes very difficult in the deformed case and so far we were
not able to solve it.

The asymmetry that appears in (59) and (60) is a consequence of the
very different expressions for the comultiplication acting on
different pairs of deformed para-Bose operators (see eq. (3.7) in
[34])). The latter have been derived from the quite symmetrical
expressions for the comultiplication defined on the Chevalley
generators [32]. We believe it will be possible to write down
new, symmetric expressions for the comultiplication and hence
for the deformed Lie-super triple generators (54). To this end
one has to use, may be, multiparametric deformations of
$U[osp(2n+1/2m)]$ as this was done for $gl(n)$ [41]. But
this is certainly another open problem.

\vskip 24pt
\noindent
{\bf Acknowledgements}
\vskip 12pt
The author is thankful to Prof. H. D. Doebner for the kind hospitality at
the Arnold Sommerfeld Institute for Mathematical Physics, where
most of the results in the present investigation have been
obtained. The constructive disscusions with Dr. N. I. Stoilova
are greatly acknowledged.

\vskip 24pt
\noindent
{\bf References}

\vskip 12pt
\settabs \+  [11] & I. Patera, T. D. Palev, Theoretical
   interpretation of the experiments on the elastic \cr

\+ [1] & Green H S 1953 {\it Phys. Rev.} {\bf 90} 270 \cr

\+ [2] & Greenberg O W and Messiah A M L 1965
         {\it Phys. Rev. B} {\bf 138 } 1155 \cr

\+ [3] & Ignatiev A Yu and Kuzmin V A 1987 {\it Yad. Fys. }
         {\bf 46} 486 [ 1987 {\it Sov. J. Nucl. Phys.} {\bf
         46}] 444  \cr

\+     & Greenberg O W and Mohapatra R N 1987 {\it Phys.
         Rev. Lett.} {\bf 59} 2507 \cr

\+ [4] & Floreanini R and Vinet L 1990 {\it J. Phys. A: Math.
         Gen.} {\bf 23} L1019 \cr

\+ [5] & Celeghini E, Palev T D and Tarlini M 1990 {\it Preprint}
         YITP/K-865 Kyoto and \cr
\+     & 1991  {\it Mod. Phys. Lett. B} {\bf 5} 187 	\cr

\+ [6] & Odaka K, Kishi T and Kamefuchi S 1991
	 {\it J. Phys. A: Math. Gen.} {\bf 24} L591 \cr

\+ [7] & Chakrabarti R and Jagannathan R 1991
         {\it J. Phys. A: Math. Gen.} {\bf 24} L711 \cr

\+ [8] & Palev T D 1993 {\it J. Phys. A} {\bf 26} L1111  \cr

\+ [9] & Palev T D 1993 {\it Lett. Math. Phys.} {\bf 28} 321 \cr

\+ [10] & Hadjiivanov L K 1993 {\it Journ. Math. Phys.} {\bf 34}
         5476 \cr

\+ [11] & Bonatsos Dennis and Daskaloyannis C 1993 {\it
          Phys. Lett.} {\bf 307B} 100 \cr

\+ [12] & Meljanac S, Milekovic M and Pallua S 1994
          Unified View of Deformed Single-mode Oscillator
	  Algebras \cr
\+      & {\it Preprint} RBI-TH-3/94 \cr

\+ [13] & Flato M, Hadjiivanov L K and Todorov I T 1993
         {\it Foundations of Physics} {\bf 23 } 571 \cr

\+ [14] & Brzezinski T, Egusquiza I L, Macfarlane A J
          1993 {\it Phys. Lett.}{\bf 311B} 202 \cr

\+ [15] & Macfarlane A J 1993  Generalized Oscillator
          Systems and Their Parabosonic Interpretation 	\cr
\+   	& {\it Preprint} DAMPT 93-37 \cr

\+ [16] & Macfarlane A J 1994 {\it  Journ. Math. Phys.} {\bf 35}
          1054  \cr

\+ [17] & Cho K H, Chaiho Rim, Soh D S and Park S U 1994
          {\it J. Phys. A: Math. Gen.} {\bf 27} 2811 \cr

\+ [18] & Chakrabarti R and Jagannathan R 1994
         {\it J. Phys. A: Math. Gen.} {\bf 27} L277 \cr

\+ [19] & Beckers J and Debergh N 1991
          {\it J. Phys. A: Math. Gen.} {\bf 247} L1277 \cr

\+ [20] & Kamefuchi S and Takahashi Y 1960 {\it Nucl. Phys.}
          {\bf 36} 177 \cr

\+ [21] & Ryan  C and Sudarshan E C G 1963 {\it Nucl. Phys.}
          {\bf 47} 207 \cr

\+ [22] & Omote M, Ohnuki Y and Kamefuchi S 1976 {\it
          Prog. Theor. Phys.} {\bf 56} 1948  \cr

\+ [23] & Ganchev A and Palev T D 1978 {\it Preprint} JINR P2-11941;
          1980 {\it J. Math. Phys.} {\bf 21} 797 \cr

\+ [24] & Kac V G 1977 {\it Adv. Math.} {\bf 26} 8 \cr

\+ [25] & Okubo S 1993 {\it University of Rochester preprint}
          ER-1335, ERO40685-784 \cr

\+ [26] & Palev T D 1982 {\it J. Math. Phys.} {\bf 23} 1100 \cr

\+ [27] & Gel'gand I M and Zetlin M L 1950 {\it Dokl.
          Akad. Nauk SSSR} {\bf 71} 1071 \cr

\+ [28] & Palev T D 1975 {\it Ann. Inst. Henri Poincar\'e}
          {\bf XXIII} 49 \cr

\+ [29] & Kac V G 1978 {\it Lect. Notes in Math.} {\bf 626}
          597 \cr

\+ [30] & Ky Nguyen Ahn, Palev T D and Stoilova N I 1982
          {\it J. Math. Phys.} {\bf 33} 1841 \cr

\+ [31] & Scheunert M 1979 {\it Lect. Notes in Math.} {\bf 716}
          (Berlin: Springer)\cr

\+ [32] & Floreanini R, Spiridonov V P and Vinet L 1991
          {\it Commun. Math. Phys.}{\bf 137} 149 \cr

\+ [33] & Palev T D 1993 {\it J. Math. Phys.} {\bf 34} 4872 \cr

\+ [34] & Palev T D and Stoilova N I 1993
          {\it Lett. Math. Phys.} {\bf 28} 187 \cr

\+ [35] & Palev T D 1993 {\it INRNE Preprint } INRNE-TH-93/7,
          {\it Lett. Math. Phys.} (to appear)\cr

\+ [36] & Biedenharn  L C  1989 {\it J.Phys. A} {\bf 22}  L873 \cr

\+ [37] & Macfarlane  A J  1989 {\it J.Phys. A} {\bf 22}  4581 \cr

\+ [38] & Sun  C P  and Fu  H C  1989 {\it J.Phys. A} {\bf 22} L983 \cr

\+ [39] & Hayashi T 1990 {\it Commun. Math. Phys.}
          {\bf 127} 129 \cr

\+ [40] & Mack G and Todorov I T 1969 {\it J. Math. Phys.}
          {\bf 10} 2078 \cr

\+ [41] & Sudbery A 1990 {\it J.Phys. A} {\bf 23}  L697 \cr

\end